\documentclass[12pt]{article}
\usepackage{wrapft}
\usepackage{bm}

\begin{document}

\title{Fermion Mass Hierarchy\\
in Lifshitz Type Gauge Theory}

\author{
Kunio \textsc{Kaneta}\footnote{E-mail: s08a204@shinshu-u.ac.jp}
and Yoshiharu \textsc{Kawamura}\footnote{E-mail: haru@azusa.shinshu-u.ac.jp} \\
{\it Department of Physics, Shinshu University, }\\
{\it Matsumoto 390-8621, Japan}
}

\date{
January 27, 2010}

\maketitle
\begin{abstract}
We study the origin of fermion mass hierarchy and flavor mixing in a Lifshitz type extension of the standard model
including an extra scalar field.
We show that the hierarchical structure can originate from renormalizable interactions.
In contrast to the ordinary Froggatt-Nielsen mechanism, the higher the dimension of associated operators, the heavier the fermion masses.
Tiny masses for left-handed neutrinos are obtained without introducing right-handed neutrinos.
\end{abstract}

The origin of fermion mass hierarchy and flavor mixing is one of the biggest problem in particle physics.
In the standard model (SM), the hierarchical structure originates from the texture of Yukawa couplings.
Because the Yukawa couplings are free parameters, their values should be determined by a theory beyond the SM.
Hence the structure of Yukawa couplings can give us valuable clues for exploring an underlying theory.

Recently, an exotic theory beyond the SM and/or the minimal supersymmetric SM (MSSM) has been proposed.\cite{K}
The candidate theory is a Lifshitz type extension of the SM and/or the MSSM.\footnote{
A Lifshitz type extension of gravity theory was proposed by Ho\v{r}ava.\cite{Horava1,Horava2,Horava3}
Properties of Lifshitz type field theory have been investigated in Ref.~\cite{LFT1,LFT2,LFT3,LFT4,LFT5,LFT6}.
}
This type of theory is assumed to have a fixed point with anisotropic scaling characterized 
by a dynamical critical exponent $z (> 1)$ above a high-energy scale $M_{\ell}$.
The system does not possess the relativistic invariance for $z \ne 1$. 
The Lorentz invariance is expected to emerge after the transition from $z \ne 1$ to $z = 1$ around $M_{\ell}$.\footnote{
In Ref.~\cite{Anselmi1,Anselmi2,Anselmi3}, properties and renormalizability for quantum field theories with Lorentz symmetry breaking terms 
have been studied intensively on the basis of $\lq\lq$weighted power counting".
Furthermore, extensions of the SM have been proposed for this framework.\cite{Anselmi4,Anselmi5}
}

In this letter, we study the origin of fermion mass hierarchy and flavor mixing in a Lifshitz type extension of the SM
including an extra scalar field.
We show that the hierarchical structure can originate from renormalizable interactions.

The basic idea is as follows.
The Lifshitz type theory can be renormalizable by power counting,
even though it contains higher-dimensional operators which make the theory with $z=1$ non-renormalizable.
Operators of dimensionality $4+r$ ($r>0$), $O^{(4+r)}$, become irrelevant ones $O^{(4+r)}/M_{\ell}^r$
after the transition from $z \ne 1$ to $z = 1$ (and the dimensional reduction if extra dimensions exist) around $M_{\ell}$.
The contributions from these operators are, in general, negligibly small if $M_{\ell}$ is sufficiently large.
For example, $M_{\ell}$ should be larger than $O(10^{15\sim 16})$~GeV in order to suppress proton decay processes.
Suppose that a symmetry is broken down spontaneously at a high-energy scale $M_{SB}$ larger than $M_{\ell}$
and $O^{(4+r)}$ change into $M_{SB}^{4+r-q} O^{(q)}$.
Then these operators can be relevant ones such as $(M_{SB}/M_{\ell})^{4+r-q} M_{\ell}^{4-q}O^{(q)}$ for $q \le 4$ below $M_{\ell}$.
Here, we assume that renormalizable terms including parameters with positive mass dimensions originate from a specific dynamics 
characterized by a scale $M_{\ell}$ and every parameter with mass dimension is given by a power of $M_{\ell}$.
In this case, the hierarchy among couplings related to $O^{(q)}$ can originate from 
the difference of exponents in $(M_{SB}/M_{\ell})^{4+r-q}$.
If we apply it to the Yukawa couplings, we find an interesting feature that
{\it the higher the dimension of associated original operators, the heavier the fermion masses}.\footnote{
This feature changes into an opposite one 
if $M_{SB}$ is smaller than $M_{\ell}$.}

For the sake of completeness, we explain the ordinary Froggatt-Nielsen mechanism\cite{FN} for our framework.
When we suppose that the Lifshitz type theory is an effective one derived from an underlying theory,
non-renormalizable terms can appear after integrating out superheavy fields.
That is, operators of dimensionality $D+z+p$, $O^{(D+z+p)}$, can be derived with the suppression factor $\Lambda^p$
in the Lifshitz type theory on $D+1$-dimensional space-time, where $\Lambda$ is a cut-off scale or a mass scale 
related to superheavy fields.
After a symmetry breaking at $M_{SB}$ smaller than $\Lambda$, 
$O^{(D+z+p)}$ change into $(M_{SB}/\Lambda)^{p} O^{(D+z)}$
and the hierarchy among couplings related to $O^{(D+z)}$ can originate from 
the difference of exponents in $(M_{SB}/\Lambda)^{p}$.\footnote{
Strictly speaking, an extra factor such as $(M_{\ell}/\Lambda)^{\gamma p}$ appears after the dimensional reduction and
the redefinition of fields where $\gamma$ is a constant related to $z$ and the dimensionality of extra space. See (\ref{f-L}).}
In this case, it is known that
{\it the higher the dimension of associated original operators, the lighter the fermion masses}.

First, let us explain the fermion mass hierarchy and the flavor mixing.
In the SM, the Yukawa interactions for quarks and charged leptons are given by
\begin{eqnarray}
\mathcal{L}_Y = f_{ij}^{(u)} \bar{q}_{Li} h_u {u}_{Rj}  +  f_{ij}^{(d)} \bar{q}_{Li} h {d}_{Rj}
 + f_{ij}^{(e)} \bar{l}_{Li} h {e}_{Rj} + \mbox{h.c.}~,
\label{LY}
\end{eqnarray}
where $f_{ij}^{(X)}$ ($X = u, d, e$) are the Yukawa couplings, $i, j$ are family indices $(i, j =1, 2, 3)$,
$\bar{q}_{Li}$ are Hermitian conjugates of left-handed quark doublets, $u_{Ri}$ are right-handed up type quark singlets, 
$d_{Ri}$ are right-handed down type quark singlets, $\bar{l}_{Li}$ are Hermitian conjugates of left-handed lepton doublets,
$e_{Ri}$ are right-handed electron type lepton singlets, $h$ (or $h_u \equiv i\tau_2 h^*$) is a weak Higgs doublet
and h.c. represents Hermitian conjugates of former terms.
Quark masses and charged lepton masses are
the eigenvalues of fermion mass matrices $M_X$ given by
\begin{eqnarray}
(M_u)_{ij} = f_{ij}^{(u)} \frac{v}{\sqrt{2}}~,~~ (M_d)_{ij} = f_{ij}^{(d)} \frac{v}{\sqrt{2}}~,~~ (M_e)_{ij} = f_{ij}^{(e)} \frac{v}{\sqrt{2}}~,
\label{MX}
\end{eqnarray}
where $v (= 246$GeV) is the vacuum expectation value (VEV) of neutral component $(h^0)$ of $h$.\footnote{
In the MSSM, 
$v/\sqrt{2}$ is replaced by the corresponding one, i.e., either VEV for neutral components $(h_u^0, h_d^0)$
of two Higgs doublets.}
Using unitary matrices $S_X$ and $T_X$, $M_X$ are diagonalized as 
\begin{eqnarray}
&~& S_u^{\dagger} M_u T_u = \mbox{diag}(m_u, m_c, m_t)~,~~ S_d^{\dagger} M_d T_d = \mbox{diag}(m_d, m_s, m_b)~,
\nonumber \\
&~& S_e^{\dagger} M_e T_e = \mbox{diag}(m_e, m_{\mu}, m_{\tau})~.
\label{MXdiag}
\end{eqnarray}
The quark flavor mixing is given by the Kobayashi-Maskawa matrix $V_{\tiny{\rm KM}} = S_u^{\dagger} S_d$.\cite{K&M}
Experimental data\cite{PDG} show the existence of fermion mass hierarchy and flavor mixing.
Because the flavor structure originates from the texture of Yukawa couplings and 
the Yukawa couplings are free parameters in the SM,
we need a theory beyond the SM to strip the structure of its aura of mystery.
Suppose that an underlying theory holds above $O(10^{15\sim 16})$~GeV.
Considering renormalization effects in the SM or the MSSM, 
the magnitude of each fermion mass and each entry in $V_{\tiny{\rm KM}}$ at $O(10^{16})$~GeV
can be roughly represented as
\begin{eqnarray}
&~& (m_u, m_c, m_t) \sim (\lambda^7, \lambda^4, 1) \frac{v}{\sqrt{2}}~,~~
(m_d, m_s, m_b) \sim (\lambda^6, \lambda^4, \lambda^2) \frac{v}{\sqrt{2}}~,
\nonumber \\
&~& (m_e, m_{\mu}, m_{\tau}) \sim (\lambda^7, \lambda^4, \lambda^2) \frac{v}{\sqrt{2}}
\label{Mexp}
\end{eqnarray}
and
\begin{eqnarray}
&~& (V_{\tiny{\rm KM}})_{ij} \sim 
\left(\begin{array}{ccc}
1 & \lambda & \lambda^3 \\
\lambda & 1 & \lambda^2 \\
\lambda^3 & \lambda^2 & 1 
\end {array}
\right)~,
\label{VKMexp}
\end{eqnarray}
where we use the Cabibbo angle $\lambda \equiv \sin\theta_C \sim 0.23$.\cite{C}
For $V_{\tiny{\rm KM}}$, recall the Wolfenstein parameterization.\cite{Wolf}
Our present goal is to derive the structure (\ref{Mexp}) and (\ref{VKMexp}) 
using a specific theory beyond the SM.
An exotic candidate beyond the SM is a Lifshitz type extension of the SM or the MSSM.

Next let us explain a framework of Lifshitz type field 
theory \cite{K,Horava1,Horava2,Horava3,LFT1,LFT2,LFT3,LFT4,LFT5,LFT6} briefly. 
Space-time is assumed to be factorized into a product of $3$-dimensional Euclidean space $\bm{R}^3$,
extra $n$-dimensional compact space and time $\bm{R}$,
whose coordinates are denoted by $x^{i}$ $(i = 1, 2, 3)$, $y^{k}$ $(k=1, \cdots, n)$ and $t$.
The notation $x^I$ $(I = 1, \cdots, n+3)$ is also used for the $(n+3)$-dimensional space coordinates.
The dimensions of $x^{i}$, $y^k$ and $t$ are defined as
\begin{eqnarray}
[x^i] = [y^k] = -1~,~~ [t]=-z~,
\label{dim1}
\end{eqnarray}
where $z$ is the dynamical critical exponent, which characterizes anisotropic scaling 
$x^i \to bx^{i}$, $y^k \to by^{k}$ and $t \to b^z t$ at the fixed point.
The system does not possess the relativistic invariance for $z \ne 1$ 
but it possesses spatial rotational invariance and translational invariance.
The kinetic terms for a complex scalar field $\Phi$ and a spinor field $\Psi$ are given by
\begin{eqnarray}
&~& \int dt d^3{x} d^n{y}
\left[ \left|\frac{\partial \Phi}{\partial t}\right|^2 + \bar{\Psi} i \Gamma^0 \frac{\partial}{\partial t} \Psi + \cdots \right]~,
\label{kinetic}
\end{eqnarray}
where $\Gamma^0$ corresponds to the time component of the gamma matrices and
the ellipsis stands for terms including spatial derivatives.
$\Psi$ is a spinor defined on $\bm{R}^{n+3}$ and it transforms as 
\begin{eqnarray}
&~& \Psi(x) \to \Psi' \to \Psi'(x') = S(O) \Psi(x)~, 
\label{Rotation}\\
&~& S(O) \equiv e^{-\frac{i}{4}\omega_{IJ}\Sigma^{IJ}}~,~~ \Sigma^{IJ} \equiv \frac{i}{2}\left(\Gamma^I \Gamma^J - \Gamma^J \Gamma^I \right)~,
\label{S(O)}
\end{eqnarray}
under the spatial rotation $x^I \to x'^{I} = O^I_J x^J$.
Here, the $\Gamma^I$ are gamma matrices, the $\omega_{IJ}$ are parameters related to the rotation angles $\theta^I$
with $\omega_{IJ} = - \varepsilon_{IJK} \theta^K$
and $O^I_J$ is the orthogonal matrix given by $\displaystyle{O^I_J = (e^{\omega})^I_J}$.
$S(O)$ satisfies the following relations:
\begin{eqnarray}
S^{\dagger}(O) \Gamma^I S(O) = O^I_J \Gamma^J~, ~~ S^{\dagger}(O) S(O) = {\mathcal{I}}~,
\label{S-rel}
\end{eqnarray}
where $\mathcal{I}$ is the unit matrix.
Chiral fermions on $\bm{R}^3$ are assumed to appear after the dimensional reduction, e.g., through the orbifold breaking mechanism.
The engineering dimensions of $\Phi$ and $\Psi$ are given by
\begin{eqnarray}
[\Phi] = \frac{3+n-z}{2}~,~~ [\Psi]=\frac{3+n}{2}~,
\label{dim}
\end{eqnarray}
respectively.
Then the dimension of operator $\Psi^{\dagger} \Phi^N \Psi$ is given by
\begin{eqnarray}
[\bar{\Psi} \Phi^N \Psi] = \frac{3+n-z}{2} N + 3+n~.
\label{dim-op}
\end{eqnarray}
The operator $\bar{\Psi} \Phi^N \Psi$ becomes relevant or a renormalizable term if its dimension is less than or equals to $3+n+z$.
The operator including higher spatial derivatives such as $\Phi^{\dagger} \nabla^{2z} \Phi$ can also become relevant.
The theory can be renormalizable by power counting,
even though it contains higher-dimensional operators which make the theory with $z=1$ non-renormalizable.
The Lorentz invariance is expected to emerge as an accidental symmetry
after the transition from a high-energy theory with $z \ne 1$ to that with $z=1$ around $M_{\ell}$.\footnote{
There has been a proposal that the Lorentz invariance appears at an attractive infra-red fixed point.\cite{NN,CN}
}
The magnitude of Lorentz symmetry breaking terms is estimated and it gives constraints 
on parameters.\cite{Lvio-test1,Lvio-test2,Lvio-test3,C&G},\cite{K}

Now let us explore the origin of texture of Yukawa couplings in a Lifshitz type extension of the SM
including an extra scalar field $\Phi$.\footnote{
The study based on the Lifshitz type extension of the MSSM can be carried out by the introduction of two Higgs doublets
and similar results can be obtained.
}
We introduce an extra $U(1)$ symmetry denoted by $U(1)_A$ and assume that
$\Phi$ is a singlet under the SM gauge group $SU(3)_C \times SU(2)_L \times U(1)_Y$ 
but has a non-zero $U(1)_A$ charge and $U(1)_A$ is spontaneously broken down by 
the non-vanishing VEV of $\Phi$ above $M_{\ell}$.
The interactions such as $\Phi^{m_{ij}^{(X)}} \bar{\Psi}_i H_{(u)} \Psi_j$ are determined by $U(1)_A$.
Here, $m_{ij}^{(X)}$ are zero or positive integers, 
and $\Psi_i$ and $H_{(u)}$ are fermions and a boson which contain SM fermions $\psi_i$
and a weak Higgs boson $h_{(u)}$ as zero modes, respectively.
The origin of extra $U(1)$ symmetry is not specified in our analysis for simplicity.\footnote{
If $U(1)_A$ is an anomalous gauge symmetry, 
we need to introduce extra fields in order for the theory to be harmless.
}
The action is given by
\begin{eqnarray}
\hspace{-1.3cm} &~& \int dt d^3{x} d^n{y}
\left[ \left|D_t \Phi\right|^2 - \frac{1}{\kappa^2} \Phi^{\dagger} (D_I^{\dagger} D_I)^z \Phi - C_{\Phi}\left|D_I \Phi\right|^2 \right.
\nonumber \\
\hspace{-1.3cm} &~& ~~~ + \left|D_t H\right|^2 - \frac{1}{\kappa^2_h} H^{\dagger} (D_I^{\dagger} D_I)^z H - C_{H}\left|D_I H\right|^2
+ \bar{\Psi}_i i \Gamma^0 D_t \Psi_i 
\nonumber \\
\hspace{-1.3cm} &~& ~~~ \left. -\frac{1}{\xi^2_i} \bar{\Psi}_i (i \Gamma^I D_I)^z \Psi_i  - C_{\Psi_i} \bar{\Psi}_i i \Gamma^I D_I \Psi_i
 + \gamma_{ij}^{(X)} \Phi^{m_{ij}^{(X)}} \bar{\Psi}_i H_{(u)} \Psi_j+ \cdots \right]~,
\label{S}
\end{eqnarray}
where $\kappa^2$, $\kappa^2_h$ and $\xi^2_i$ are dimensionless parameters concerning Lorentz symmetry violating terms,
$D_t$ and $D_I$ are covariant derivatives and the ellipsis stands for other terms.
The engineering dimensions of fields $(\Phi, H_{(u)}, \Psi_i)$ and parameters $(C_{\Phi}, C_{H}, C_{\Psi_i}, \gamma_{ij}^{(X)})$ are given by
\begin{eqnarray}
[\Phi] = [H_{(u)}] = \frac{3+n-z}{2}~,~~ [\Psi_i]=\frac{3+n}{2}
\label{dim-fields}
\end{eqnarray}
and
\begin{eqnarray}
&~& [C_{\Phi}] = [C_H]= 2(z-1)~,~ [C_{\Psi_i}] = z-1~,~ \nonumber \\
&~& [\gamma_{ij}^{(X)}]=z-\frac{(3+n-z)(m_{ij}^{(X)}+1)}{2}~,
\label{dim-parameters}
\end{eqnarray}
respectively.
We assume that renormalizable terms including parameters with positive mass dimensions originate from a specific dynamics 
characterized by a scale $M_{\ell}$ and the parameters are given by a power of $M_{\ell}$,
which is similar to the soft supersymmetry (SUSY) breaking parameters in SUSY models.\footnote{
For example, the soft supersymmetry breaking terms are given by a power of the gravitino mass $m_{3/2}$ in the gravity mediation.}
The magnitude of parameters is not necessarily $O(1)$ in the unit of $M_{\ell}$
but can be much smaller like most $A$ terms in SUSY models.
On the other hand, we assume that parameters in non-renormalizable terms are suppressed by a power of cutoff scale $\Lambda$
as usual.
To become relativistic below $M_{\ell}$,
finetuning among parameters is required such as $C_{\Phi} = C_H = C_{\Psi_i}^2$ for all species.
In this setup, parameters are expressed as
\begin{eqnarray}
&~& C_{\Phi}= C_H = M_{\ell}^{2(z-1)}~,~~ C_{\Psi_i} = M_{\ell}^{z-1}~,\nonumber \\
&~& \gamma_{ij}^{(X)} = \left\{
\begin{array}{ll}
\displaystyle{\gamma_{ij}^{0(X)} M_{\ell}^{z-\frac{(3+n-z)(m_{ij}^{(X)}+1)}{2}}} &
~~\left(z \geq \frac{(3+n-z)(m_{ij}^{(X)}+1)}{2}\right)~,\\
\displaystyle{\gamma_{ij}^{0(X)} \Lambda^{z-\frac{(3+n-z)(m_{ij}^{(X)}+1)}{2}}}&
~~\left(z < \frac{(3+n-z)(m_{ij}^{(X)}+1)}{2}\right)~,
\end{array}\right.
\label{para}
\end{eqnarray}
after a suitable rescaling of fields.
Here, $\gamma_{ij}^{0(X)}$ is a dimensionless parameter.
We assume that the volume of extra $n$-dimensional space is $1/M_{\ell}^n$.
After the redefinition of time variable and fields as
\begin{eqnarray}
&~& x_0 \equiv M_{\ell}^{z-1} t~,~~ \tilde{\Phi} \equiv M_{\ell}^{\frac{z-n-1}{2}} \Phi = \phi + \cdots~,~~ 
\nonumber \\
&~& \tilde{H}_{(u)} \equiv M_{\ell}^{\frac{z-n-1}{2}} H_{(u)} = h_{(u)} + \cdots~,~~ 
\tilde{\Psi}_i \equiv M_{\ell}^{-\frac{n}{2}} \Psi_i = \psi_i + \cdots
\label{redef}
\end{eqnarray}
and the dimensional reduction of extra dimensions,
the following action for zero modes is derived from (\ref{S}),
\begin{eqnarray}
\int d^4{x} 
\left[ \left|D_{\mu} h\right|^2 + \bar{\psi}_i i \gamma^{\mu} D_{\mu} \psi_i 
 + \gamma_{ij}^{0(X)} \left(\frac{\langle \phi \rangle}{M_{\ell}}\right)^{m_{ij}^{(X)}} \bar{\psi}_i {h}_{(u)} \psi_j + \cdots \right]~,
\label{S4D}
\end{eqnarray}
where the ellipsis  stands for the Yukawa interactions from non-renormalizable terms and so on.
The dimensions of $\phi$, $h_{(u)}$ and $\psi_i$ are $[\phi] = [h_{(u)}] = 1$ and $[\psi_i]=3/2$.
The Yukawa couplings are given by\footnote{
The contributions from volume suppression factor can also appear after the dimensional reduction.\cite{Y}
In this case, the difference of field configurations related to interactions can be important to study the origin of mass hierarchy.
Here, we do not consider them for simplicity.
}
\begin{eqnarray}
f_{ij}^{(X)} = \left\{
\begin{array}{ll}
\displaystyle{\gamma_{ij}^{0(X)} \left(\frac{\langle \phi \rangle}{M_{\ell}}\right)^{m_{ij}^{(X)}}}  &
\left(z \geq \frac{(3+n-z)(m_{ij}^{(X)}+1)}{2}\right)~,\\
\displaystyle{\gamma_{ij}^{0(X)} \left(\frac{M_{\ell}}{\Lambda}\right)^{\frac{3+n-3z}{2}} 
\left(\left(\frac{M_{\ell}}{\Lambda}\right)^{\frac{1+n-z}{2}}\frac{\langle \phi \rangle}{\Lambda}\right)^{m_{ij}^{(X)}}} &
\left(z < \frac{(3+n-z)(m_{ij}^{(X)}+1)}{2}\right)~.
\end{array}\right.
\label{f-L}
\end{eqnarray}
The exponents $m_{ij}^{(X)}$ are determined from the $U(1)_A$ charge conservation:
\begin{eqnarray}
&~& m_{ij}^{(u)} Q_A(\phi) + Q_A(\bar{q}_{Li}) + Q_A(u_{Rj}) + Q_A(h_u) = 0~,
\label{QAu}\\
&~& m_{ij}^{(d)} Q_A(\phi) + Q_A(\bar{q}_{Li}) + Q_A(d_{Rj}) + Q_A(h) = 0~,
\label{QAd}\\
&~& m_{ij}^{(e)} Q_A(\phi) + Q_A(\bar{l}_{Li}) + Q_A(e_{Rj}) + Q_A(h) = 0~,
\label{QAe}
\end{eqnarray}
where $Q_A$ represents the charge of $U(1)_A$.
The first one in (\ref{f-L}) comes from renormalizable terms  
and the ratio ${M}_{\ell}/{\langle \phi \rangle}$ can play a role of $\lambda$,
in the case that there is no hierarchy among each entry in $\gamma_{ij}^{0(X)}$
and $\langle \phi \rangle$ is larger than $M_{\ell}$.
In other words, there is a build-in mechanism to generate the hierarchy using the ratio ${M}_{\ell}/{\langle \phi \rangle}$
on the basis of relevant operators.
On the other hand, the second one in (\ref{f-L}) comes from 
non-renormalizable terms which might originate from
some renormalizable interactions after integrating out superheavy fields.
The mechanism to generate the hierarchy using the ratio 
$\displaystyle{\left({M_{\ell}}/{\Lambda}\right)^{\frac{1+n-z}{2}}{\langle \phi \rangle}/{\Lambda}}$
is regarded as the Lifshitz type extended version of the Froggatt-Nielsen mechanism.\footnote{
We need some selection rule related to interactions in order for the Froggatt-Nielsen mechanism to work.
In most cases, one uses an anomalous $U(1)$ gauge symmetry,\cite{An1,An2} whose anomalies are canceled 
via the Green-Schwarz mechanism,\cite{G&S}
motivated by superstring theories.
A discrete horizontal symmetry is also used.\cite{dis}
}
There is a possibility that the hierarchy of Yukawa couplings stems from the mixture of first and second ones.

Next we consider the case with $z =4$ and $n=2$ as an example.
The Yukawa couplings are given by 
\begin{eqnarray}
f_{ij}^{(X)} = \left\{
\begin{array}{ll}
\displaystyle{\gamma_{ij}^{0(X)} \left(\frac{\langle \phi \rangle}{M_{\ell}}\right)^{m_{ij}^{(X)}}}  &
(0 \leq m_{ij}^{(X)} \leq 7)~,\\
\displaystyle{\gamma_{ij}^{0(X)} \left(\frac{\Lambda}{M_{\ell}}\right)^{\frac{7}{2}} 
\left(\frac{\langle \phi \rangle}{\sqrt{M_{\ell}\Lambda}}\right)^{m_{ij}^{(X)}}} &
(m_{ij}^{(X)} >7)~.
\end{array}\right.
\label{f-Lz=4}
\end{eqnarray}
If ${M}_{\ell}/{\langle \phi \rangle} \sim \lambda$, the fermion mass hierarchy (\ref{Mexp}) can be derived 
from the $U(1)_A$ charge assignment:\footnote{
The $U(1)_A$ charge assignment is not unique.}
\begin{eqnarray}
&~& Q_A(\bar{q}_{Li}) = (0, 1, 3)~,~~ Q_A(u_{Ri}) = (0, 2, 4)~,~~ Q_A(d_{Ri}) = (1, 2, 2)~,
\nonumber \\
&~& Q_A(\bar{l}_{Li}) = (0, 1, 1)~,~~ Q_A(e_{Ri}) = (0, 2, 4)~,~~ Q_A(h_{(u)})=0~,~~ 
\nonumber \\
&~& Q_A(\phi)= -1~,
\label{QAOurs}
\end{eqnarray}
where we assume $\displaystyle{\gamma_{ij}^{0(X)} = O((M_{\ell}/\langle \phi \rangle)^7)}$.
If ${\langle \phi \rangle}/{\sqrt{M_{\ell} \Lambda}} \sim \lambda$,
the fermion mass hierarchy (\ref{Mexp}) can be derived from the $U(1)_A$ charge assignment:
\begin{eqnarray}
&~& Q_A(\bar{q}_{Li}) = (11, 10, 8)~,~~ Q_A(u_{Ri}) = (4, 2, 0)~,~~ Q_A(d_{Ri}) = (3, 2, 2)~,
\nonumber \\
&~& Q_A(\bar{l}_{Li}) = (9, 8, 8)~,~~ Q_A(e_{Ri}) = (6, 4, 2)~,~~ Q_A(h_{(u)})=0~,~~ 
\nonumber \\
&~& Q_A(\phi)= -1~,
\label{QAFN}
\end{eqnarray}
where we assume $\displaystyle{\gamma_{ij}^{0(X)} = O(\Lambda^{1/2} M_{\ell}^{15/2}/{\langle \phi \rangle}^8)}$.
In either case, the quark flavor mixing due to the Kobayashi-Maskawa matrix (\ref{VKMexp}) can be obtained by 
\begin{eqnarray}
&~& (V_{\tiny{\rm KM}})_{ij} \sim \lambda^{|Q_A(\bar{q}_{Li})-Q_A(\bar{q}_{Lj})|} \sim 
\left(\begin{array}{ccc}
1 & \lambda & \lambda^3 \\
\lambda & 1 & \lambda^2 \\
\lambda^3 & \lambda^2 & 1 
\end {array}
\right)~.
\label{VKMFN}
\end{eqnarray}

Finally, we discuss the neutrino sector.
The Majorana mass matrix $M_{\nu}$ of left-handed neutrinos is usually obtained throught the see-saw mechanism
such that\cite{see-saw1,see-saw2,see-saw3}
\begin{eqnarray}
(M_{\nu})_{ij} = f_{ia}^{(\nu)} \frac{v}{\sqrt{2}} \left(M_R^{-1}\right)_{ab} f_{bj}^{(\nu)} \frac{v}{\sqrt{2}}~,
\label{Mmu}
\end{eqnarray}
where $f_{ia}^{(\nu)}$ is the Yukawa coupling among $l_{Li}$, right-handed neutrinos $\nu_{Ra}$ and $h_u$
and $(M_R)_{ab}$ is the superheavy Majorana mass matrix of right-handed neutrinos.
In our Lifshitz type extension of the SM including $\Phi$, $M_{\nu}$ can be obtained 
without introducing right-handed neutrinos from the following relevant interactions:
\begin{eqnarray}
\gamma_{ij}^{(\nu)}  {\Phi}^{m_{ij}^{(\nu)}} \bar{L}_{i} i\tau_2 {\bm{\tau}} L_{j}^{c} \cdot H_u^t i\tau_2 {\bm{\tau}} H_u~,
\label{MmuInt}
\end{eqnarray}
where $L_{j}$ are fermions whose zero modes are $l_{Li}$, and
superscripts $c$ and $t$ stand for the charge conjugation and transpose of the relevant field, respectively.
The exponent $m_{ij}^{(\nu)}$ is determined by
\begin{eqnarray}
&~& m_{ij}^{(\nu)} Q_A(\phi) + Q_A(\bar{l}_{Li}) + Q_A(\bar{l}_{Lj}) + 2 Q_A(h_u) = 0~.
\label{QAnu}
\end{eqnarray}
In our model, $M_{\nu}$ is given by
\begin{eqnarray}
\hspace{-0.3cm} (M_\nu)_{ij} = \left\{
\begin{array}{ll}
\!\!\! \displaystyle{\gamma_{ij}^{0(\nu)}\!\! \left(\frac{\langle \phi \rangle}{M_{\ell}}\right)^{m_{ij}^{(\nu)}}\!\!\!\!\!
\frac{v^2}{2M_{\ell}}}  &
\!\!\! \left(z \geq \frac{(3+n-z)(m_{ij}^{(\nu)}+2)}{2}\right)~,\\
\!\!\! \displaystyle{\gamma_{ij}^{0(\nu)}\!\! \left(\frac{M_{\ell}}{\Lambda}\right)^{2+n-2z}\!\!\!
\left(\left(\frac{M_{\ell}}{\Lambda}\right)^{\frac{1+n-z}{2}}\frac{\langle \phi \rangle}{\Lambda}\right)^{m_{ij}^{(\nu)}}\!\!\!\!\!
\frac{v^2}{2\Lambda}} &
\!\!\! \left(z < \frac{(3+n-z)(m_{ij}^{(\nu)}+2)}{2}\right)~,
\end{array}\right.
\label{Mmu-L}
\end{eqnarray}
where $\gamma_{ij}^{0(\nu)}$ is a dimensionless parameter.
Using a unitary matrix $S_{\nu}$, $M_{\nu}$ is diagonarized as
\begin{eqnarray}
S_{\nu}^{\dagger} M_{\nu} S_{\nu} = \mbox{diag}(m_{\nu_1}, m_{\nu_2}, m_{\nu_3})~,
\label{Mnudiag}
\end{eqnarray}
where $m_{\nu_1} < m_{\nu_2} < m_{\nu_3}$.
The lepton flavor mixing is given by the Maki-Nakagawa-Sakata matrix $V_{\tiny{\rm MNS}} = S_e^{\dagger} S_{\nu}$.\cite{MNS}
Using the experimental data for neutrinos,\cite{PDG} we find that there are two large mixings:
\begin{eqnarray}
\sin^2 2\theta_{12} \sim 0.88~,~~ \sin^2 2\theta_{23} > 0.92
\label{bi-large}
\end{eqnarray}
and the hierarchy between mass-squared differences for solar neutrinos $\Delta m_{\odot}^2$ 
and for the atmospheric neutrinos $\Delta m_{\oplus}^2$:
\begin{eqnarray}
\frac{\Delta m_{\odot}^2}{\Delta m_{\oplus}^2} = \frac{|m_{\nu_2}^2 - m_{\nu_1}^2|}{|m_{\nu_3}^2 - m_{\nu_2}^2|} \sim \lambda^2~.
\label{DeltaM}
\end{eqnarray}
If three neutrino masses do not degenerate, the hierarchy (\ref{DeltaM}) suggests the relation:
\begin{eqnarray}
(m_{\nu_2}, m_{\nu_3}) \sim (\lambda, 1) \times 0.05 \mbox{eV}~.
\label{Mnuexp}
\end{eqnarray}
Using our mechanism with the $U(1)$ charge assignment (\ref{QAOurs})
or the ordinary Froggatt-Nielsen mechanism with the $U(1)$ charge assignment (\ref{QAFN}),
$M_{\nu}$ and $V_{\tiny{\rm MNS}}$ are estimated as
\begin{eqnarray}
&~& (M_{\nu})_{ij} \propto \lambda^{\mp(Q_A(\bar{l}_{Li})+Q_A(\bar{l}_{Lj}))} \propto
\left(\begin{array}{ccc}
\lambda^2 & \lambda & \lambda \\
\lambda & 1 & 1 \\
\lambda & 1 & 1 
\end {array}
\right)~,
\label{MnuFN}\\
&~& V_{\tiny{\rm MNS}} \sim \lambda^{|Q_A(\bar{l}_{Li})-Q_A(\bar{l}_{Lj})|} S_{\nu} \sim
\left(\begin{array}{ccc}
1 & \lambda & \lambda \\
\lambda & 1 & 1 \\
\lambda & 1 & 1 
\end {array}
\right) S_{\nu}~,
\label{VMNSFN}
\end{eqnarray}
where the minus and the plus sign in (\ref{MnuFN}) for our mechanism and the Froggatt-Nielsen mechanism, respectively.
This type of neutrino mass matrix has been proposed and studied in Refs.~\cite{S&Y,V,L&R}.
The above matrices (\ref{MnuFN}) and (\ref{VMNSFN}) have an interesting property
that the bi-large mixing can be naturally derived if we obtain the mass relation (\ref{Mnuexp}) after the diagonalization of $M_{\nu}$.

Finally, we discuss the lepton number and/or baryon number violating process through four fermi interactions.
The four fermi interactions originate from the operator such as $\Phi^N \Psi^{\dagger} \Psi \Psi^{\dagger} \Psi$ 
after $\Phi$ acquires the VEV, and there appear the lepton number and/or baryon number violating terms such as
$qqql$ and $udde$.
The dimension of $\Phi^N \Psi^{\dagger} \Psi \Psi^{\dagger} \Psi$ is given by
\begin{eqnarray}
[\Phi^N \Psi^{\dagger}  \Psi \Psi^{\dagger}  \Psi] = \left(\frac{N}{2} + 1\right) (3+n-z) + 3 + n + z~.
\label{dim-fourfermi}
\end{eqnarray} 
The operator $\Phi^N \Psi^{\dagger}  \Psi \Psi^{\dagger}  \Psi$ becomes a non-renormalizable term by a power-counting
if $z$ is less than $3+n$, and then the interaction is suppressed by a power of $\Lambda$.

The origin of fermion mass hierarchy and flavor mixing has been studied using the Froggatt-Nielsen mechanism
in the framework of (supersymmetric) grand unified theory.\cite{S&Y,N&Y,B&K1,B&K2,M}
A similar analysis can be carried out in the framework of Lifshitz type extension of (supersymmetric) grand unified theory.

In conclusion, we have studied the origin of fermion mass hierarchy and flavor mixing in a Lifshitz type extension of the SM
including an extra scalar field.
We have found a mechanism to generate the hierarchical structure from renormalizable interactions.
The mechanism is similar to the Froggatt-Nielsen mechanism in the sense that the hierarchy originates from
operators of dimensionality $u$ $(u > 4)$.
But, there is a difference in characters between them.
In the ordinary Froggatt-Nielsen mechanism, the higher the dimension of associated operators, the lighter the fermion masses.
In our mechanism, the higher the dimension of associated operators, the heavier the fermion masses.
Furthermore we have found that tiny masses for left-handed neutrinos can be obtained without introducing right-handed neutrinos.

\section*{Acknowledgements}
This work was supported in part by scientific grants from the Ministry of Education, Culture,
Sports, Science and Technology under Grant Nos.~18204024 and 18540259 (Y.K.).

\end{document}